\documentclass{article}
\pdfoutput=1

\usepackage{graphicx}
\usepackage{xcolor}
\usepackage{wrapfig}
\usepackage[ruled,vlined]{algorithm2e}

\usepackage[preprint, nonatbib]{neurips_2020}

\usepackage[utf8]{inputenc} 
\usepackage[T1]{fontenc}    
\usepackage{hyperref}       
\usepackage{url}            
\usepackage{booktabs}       
\usepackage{amsfonts}       
\usepackage{nicefrac}       
\usepackage{microtype}      

\usepackage[numbers]{natbib}
\usepackage{amsmath, amssymb}
\usepackage[font=small]{caption, subcaption}

\newcommand{\bz}{\mathbf{z}}

\newcommand{\bx}{\mathbf{x}}
\newcommand{\bw}{\mathbf{w}}
\newcommand{\bff}{\mathbf{f}}
\newcommand{\bu}{\mathbf{u}}

\newcommand{\bphi}{\boldsymbol{\varphi}}

\newcommand{\GalerkinBasisModel}{\texttt{GalerkinBasis}}
\newcommand{\ConvectionModel}{\texttt{Convection}}
\newcommand{\MyModel}{\texttt{DOD-Convection}}

\title{Deep Orthogonal Decompositions for Convective Nowcasting}

\author{%
  Daniel J. Tait
  \\
  University of Warwick \\
  Alan Turing Institute \\
  \texttt{dtait@turing.ac.uk} \\
}

\begin{document}

\maketitle

\begin{abstract}
Near-term prediction of the structured spatio-temporal processes driving our climate is of profound importance to the safety and well-being of millions, but the prounced nonlinear convection of these processes make a complete mechanistic description even of the short-term dynamics challenging. However, convective transport provides not only a principled physical description of the problem, but is also indicative of the transport in time of informative features which has lead to the recent successful development of ``physics free'' approaches to the now-casting problem. In this work we demonstrate that their remains an important role to be played by physically informed models, which can successfully leverage deep learning (DL) to project the process onto a lower dimensional space on which a minimal dynamical description holds. Our approach synthesises the feature extraction capabilities of DL with physically motivated dynamics to outperform existing model free approaches, as well as state of the art hybrid approaches, on complex real world datasets including sea surface temperature and precipitation.
\end{abstract}

\section{Introduction}
Nowcasting refers to the process of producing near-term predictions 
of the complex dynamical systems 
driving our natural world. Such systems have a significant 
impact on the daily lives and safety of our planets' population,
with important examples including precipitation and sea surface 
temperature; variables crucial for flood warning and 
prevention, prediction of cyclogenesis, and maritime safety. 
A common theme is the dominance of short-term dynamics by convection
-- the transport of material along the stream lines of a vector 
field. Transport and the reduced role of diffusion at these 
time-scales leads to temporal feature preservation, and 
consequently these systems have received increasing interest from 
the machine learning (ML) community 
\citep{shi, shi15, DBLP:journals/corr/MathieuCL15, vondrick, google}.

The classical approach to solving these problems in the natural
sciences is to specify a model of the underling physical properties. A 
properly calibrated mechanistic model described by parameterised 
differential operators is able to provide future prediction and 
generalise to new scenarios, provided this change of 
physics can be encoded by some parameter. In a lingua 
franca of ML we would comment on the remarkable 
generalisation and transfer-ability of fundamental physical theories 
specified by mathematical models, a demonstration of ``the unreasonable 
effectiveness of mathematics'' \citep{wigner1960}.

Nevertheless, a successful physical theory often requires multiple 
simplifying assumptions so removing many of the 
complexities of real-world phenomena, done correctly this 
reduction allows the theory to focus on the most salient 
qualities of the system. However, this reduction can be
insufficient for detailed prediction in the most complex open world 
systems we experience in nature, or else require
expensive numerical models \citep{madec}. This complexity, combined with the
existence of the persistent, informative features that characterise
convection has led to the development of deep 
learning (DL) approaches, notably the dynamics based 
approaches of \citep{shi, shi15, DBLP:journals/corr/MathieuCL15, vondrick}, 
or the images-to-images type approaches adopted by 
\citep{AYZEL2019186, lebedev, google}. As DL methods continue to advance 
it becomes an open question as 
to whether one should still attempt to embed mechanistic structure 
into our models, or if one ought to go ``physics free'' \citep{google}. 
We suggest this stance is information 
inefficient, ignoring the vast prior scientific knowledge acquired by human
endeavour. This viewpoint has led 
to the emergence of \emph{physics informed} ML 
\citep{Raissi, de2018deep} which aims to combine the 
expressive power of modern ML methods, with the representative power 
of fundamental physical models. Typically these employ DL surrogates of the 
input-output map, then ensuring this
map is either constructed to obey some physical principle
\citep{pmlr-v97-cohen19d, NIPS2019_9580}, or else 
regularised by a guiding mechanistic model 
\citep{Raissi, SIRIGNANO20181339, Berg, RAISSI2019686, Yazdani865063, zhu}

In this work we add to the argument that physical principles still 
have an important
role to play in modelling by introducing a new physically informed method 
motivated by two of the guiding principles behind a successful mathematical
analysis of a complex dynamical problem; (i) the
projection of a high dimensional model onto a lower dimensional representation, 
and (ii) the existence of a transformation that brings the dynamics 
closer to linear. 
We use the feature extraction prowess
of DL to perform our reduction, while using physical
principles to guide the evolution of these features,
constructing a flexible system that is well suited to modelling
convection phenomena by ensuring that deep nonlinear components are strongly 
linked with the minimal physical dynamics.

In summary, we use the powerful feature extraction capabilities of modern 
ML to perform localised model order reduction by projection onto a 
subspace spanned by deeply-extracted orthogonal features. We then show 
how one can combine flexible methods with minimal physical 
representations to advance the dynamics in this reduced space,
crucially we construct our method so as to enable a tighter
link between the DL-based encoding 
and the dynamics than previous approaches.
Finally we demonstrate that this synthesis of DL feature 
extraction and minimal physics outperforms both 
existing model free approaches and state of the art hybrid 
approaches on complex real world data-sets.

\section{Background}
\label{sec:background}
The classical approach to model based forecasting 
predicts the evolution of a variable
using past records to calibrate a mechanistic model. 
These models are typically complex systems of 
partial differential equations (PDEs), which must be 
sufficiently rich to embed all relevant physical
knowledge. These systems can be used to describe
the change in intensity of some target variable, for example 
sea surface temperature or precipitation rate. Two of the
most important physical processes describing these changes 
are diffusion -- a smoothing transformation, 
and transport -- volume
preserving translations, and
we now provide a brief review of the mathematical 
models encoding these effects.

\begin{wrapfigure}{R}{0.5\textwidth}
\begin{minipage}{\linewidth}
    \centering\captionsetup[subfigure]{justification=centering}
            \begin{align}
                \frac{\partial u}{\partial t} = 
                -\nabla \cdot (\alpha \nabla u)
                + \mathbf{w} \cdot \nabla \mathbf{u}
                \label{eq:lin_transport}
            \end{align}
        \subcaption{Pseudo-linear transport}\label{eq:pseudo_linear_transport}
        \begin{subequations}
            \begin{align}
                &\frac{\partial u}{\partial t} + \bw \cdot \nabla u = 
                    \nabla \cdot (\alpha \nabla u) 
                \label{eq:raleigh_benard_a} \\
                &\frac{\partial \bw}{\partial t} + \bw \cdot \nabla \bw = \nu \Delta \bw + f
                \label{eq:raleigh_benard_b} \\
                &\nabla \cdot \mathbf{w} = 0
                \label{eq:raleigh_benard_c}
            \end{align} \label{eq:raleigh_benard}
        \end{subequations} 
        \subcaption{Rayleigh - B\'enard convection} 
\end{minipage}
\end{wrapfigure}
\subsection{Models for transport and diffusion}
Our goal is to introduce a flexible family of models for the evolution
of a field variable $u(x, t)$ indexed by spatial coordinates
$\bx$ in a region $\Omega \subset \mathbb{R}^D$, and 
temporal coordinate $t \in [t_0, T]$. A starting point for 
describing diffusion and transport is  the PDE
\eqref{eq:lin_transport}, which is parameterised 
by a diffusion coefficient $\alpha(\bx, u)$, governing smoothing
over time, and a transport vector field, $\mathbf{w}(\bx, u)$,
describing the direction of bulk motion, this a (pseudo-)linear PDE 
and so in principle easy to solve. However, this only defers
modelling complexity because one must still motivate an 
appropriate transport vector field. Mathematically the most 
well studied models jointly describing convective
transport are the system \ref{eq:raleigh_benard} of 
Rayleigh-B\'enard equations \citep{saltzman, spiegel}.
However, solving this system across multiple spatial and
temporal scales, 
even for known parameters, presents a formidable challenge.

Nevertheless, this motivating idea of forming a conditionally 
linear PDE has been successfully applied by \citep{de2018deep}. 
Their approach uses data to inform the vector field 
$\mathbf{w}$, and then use the Gaussian integral kernel 
obtained by solving \eqref{eq:lin_transport} to motivate an 
integral-convolution update of the current image of the field variable. 
They propose to discard the non-linear components 
\eqref{eq:raleigh_benard_b} -- 
\eqref{eq:raleigh_benard_c} of the system 
\eqref{eq:raleigh_benard} when advancing the solution, 
then reincorporate them in the supervised per-time step loss 
function
\begin{align}
    \mathcal{L}_t = \sum_{x \in \Omega} 
    \rho(\hat{u}_{t+1}(x) - u_{t+1}(x)) 
    + \lambda_{\text{div}}\| \nabla \cdot \bw(x) \|^2 
    + \lambda_{\text{magn}} \| \hat{\bw}(x) \|^2 
    + \lambda_{\text{grad}} \| \nabla \bw (x) \|^2
    \label{eq:their_supervised_loss}
\end{align}
where $\rho$ is a loss function on the space of pixel-images, 
$\hat{u}_{t+1}$ is the predicted surface and $u_{t+1}$ is the 
actual surface. Ultimately, this assumes the linear PDE
\eqref{eq:raleigh_benard_a} is sufficient to capture the dynamics
in the original space, instead we shall attempt to project 
the dynamics onto a space where this assumption becomes 
more plausible, but first we briefly review how to numerically solve a 
PDE like \eqref{eq:lin_transport} or \eqref{eq:raleigh_benard}.

\subsection{Ritz-Galerkin discretisation of PDEs}
\label{sec:ritz-galerkin}
It is typically necessary to solve a PDE problem, like \eqref{eq:lin_transport}, numerically via discretisation, one powerful method of doing so is 
to first multiply the problem by a test function $v \in \hat{V}$, 
and then integrate to form a variational problem. It is 
typical to choose tests functions which vanish on the boundary, and so one 
arrives at the classical \emph{variational form} of the PDE problem: find 
$u \in V$ such that
\begin{align}
    \int_{\Omega}\frac{\partial u(t, \bx)}{\partial t} v(\bx) d\bx = 
    \int_{\Omega} \alpha(\bx, u) \nabla u(t, \bx) \cdot \nabla v(\bx) d\bx 
    + \int_{\Omega} \bw(\bx) \cdot \nabla u(t, \bx) v(\bx) d\bx 
    \label{eq:finite_var_problem}
\end{align}
for any test function $v \in \hat{V}$, this is the \emph{weak form} 
of the classical problem \eqref{eq:lin_transport}. In order 
to numerically implement this idea one must also replace the trial 
and test spaces by finite dimensional sub-spaces and we take $V = \mathrm{span}\left(\{ \varphi_m \}_{m=1}^M\right) = \hat{V}$. 
While there are multiple choices of basis functions 
$\{ \varphi_m \}_{m=1}^M$, for example one can use the nodal 
basis functions of the FEM \citep{reddy}, in this work we shall 
consider sets of orthonormal basis functions. That is we specify a collection of $M$ basis functions such that 
$\langle \varphi_i , \varphi_j \rangle_{L^2(\Omega)} = \delta_{ij}$, 
where $\langle \cdot, \cdot \rangle_{L^2(\Omega)}$ is the 
$L^2$-inner product.We then search for solutions 
to \eqref{eq:finite_var_problem} with representation 
$u(t, \bx) = \sum_{m=1}^M (\bz_t)_{m} \varphi_m(\bx)$, where 
$\bz_t \in \mathbb{R}^{M}$ is a vector of unknown coefficients to be determined.
Inserting this representation into \eqref{eq:finite_var_problem} we achieve a 
finite-dimensional projection of the dynamical problem as an ordinary 
differential equation (ODE)
\begin{align}
    \frac{d}{dt} \mathbf{z} = \mathbf{L}_{\varphi} \mathbf{z},
    \qquad (\mathbf{L}_{\varphi})_{ij} = \int_{\Omega} \alpha(\bx) \nabla \varphi_j(\bx)
\cdot \nabla \varphi_i(\bx) dx 
+ \int_{\Omega} \boldsymbol{\tau}(\bx) \cdot \nabla \varphi_j(\bx) 
\varphi_i (\bx)dx
    \label{eq:disc_lin_ode}
\end{align}
This is the Ritz-Galerkin projection of the dynamics onto the
subspace $V$. We use the notation $\mathbf{L}_{\varphi}$ to make it 
clear that 
where as the classical operator in the RHS of \eqref{eq:lin_transport} 
depended only up on the coefficient functions parameterising it, the projected
problem further depends on the ability of the basis functions to capture 
information about the problem, entwining the encoding and the dynamics. The 
less informative the basis functions the higher the dimension of $M$ will 
be needed to faithfully reproduce the dynamics in the function space. 
In what follows we refer to the process of forming the state dynamic 
matrix $\mathbf{L}$ in \eqref{eq:disc_lin_ode} as the \texttt{Assembly} 
operation, which involves performing quadratures to compress the 
parameter fields $\alpha, \bw$ and basis functions into an 
$M \times M$ matrix.

\subsection{Proper orthogonal decompositions}
\label{sec:pod}
Given an ensemble $\{u_t\}_{t=1}^N$ of field variables over
a domain $\Omega$, the \emph{proper orthogonal decomposition}
(POD) is a technique for extracting an informative modal basis,
that is a low-dimensional basis which captures most of the
information or ``energy'' of the ensemble. The decomposition 
in POD is more familiar to the machine learning 
community under the name principal components analysis (PCA), 
or alternatively the Karhunen-Loeve decomposition 
\citep{karhunen, loeve}, and involves reconstructing elements
of the ensemble as $\hat{u} = \sum_m \bz_k \varphi_m(\bx)$
where $\{ \varphi_m(\bx) \}_{m=1}^M$ are the first $M$-eigenfunctions
of the empirical covariance matrix of the ensemble ordered
by largest eigenvalue. 

The idea of using the POD eigenfunctions, as an ``empircal basis'' onto 
which to perform the above Ritz-Galerkin projection for modelling turbulent 
flows was introduced in \citep{LUMLEY1981215}. However, the 
optimality results 
concern POD as a linear re-constructor,
and do not transfer to any optimality on the dynamic prediction problem
\citep{berkooz}. Therefore, in this work we shall instead use deep networks 
to extract our
orthogonal subspace, however motivated by the POD idea we shall attempt to still
include some version of this optimal reconstruction to motivate the
regulariser in our supervised loss.
\section{Methodology}
\label{sec:methodology}
Our objective will be to estimate a future length $T$ sequence 
of realisations of the process starting from time $t$, using 
only information coming from the length $\ell$ history process $\{u_{k}(\bx) \}_{k=t-\ell}^{t}$. Infact, we shall
also discretise the domain as an $n_{x} \times n_y$ pixelated grid,
and instead aim to estimate the vectorised field variable $\mathbf{u} \in
\mathbb{R}^{n_x \times n_y}$. That is we aim to learn an images-to-images
map $\{ \hat{\bu}_{k} \}_{k={t+1}}^{k={t+T}} = f(\bu_{t-\ell}, \ldots, \bu_{t})$, 
which also embodies some minimal dynamical representation. To do so we introduce 
a DL approach to building a localised version of the POD-basis discussed above.

\subsection*{Deep Galerkin Features}
\begin{figure}[t!]
    \centering
    \begin{subfigure}[t]{0.6\textwidth}
        \centering
        \includegraphics[width=\linewidth]{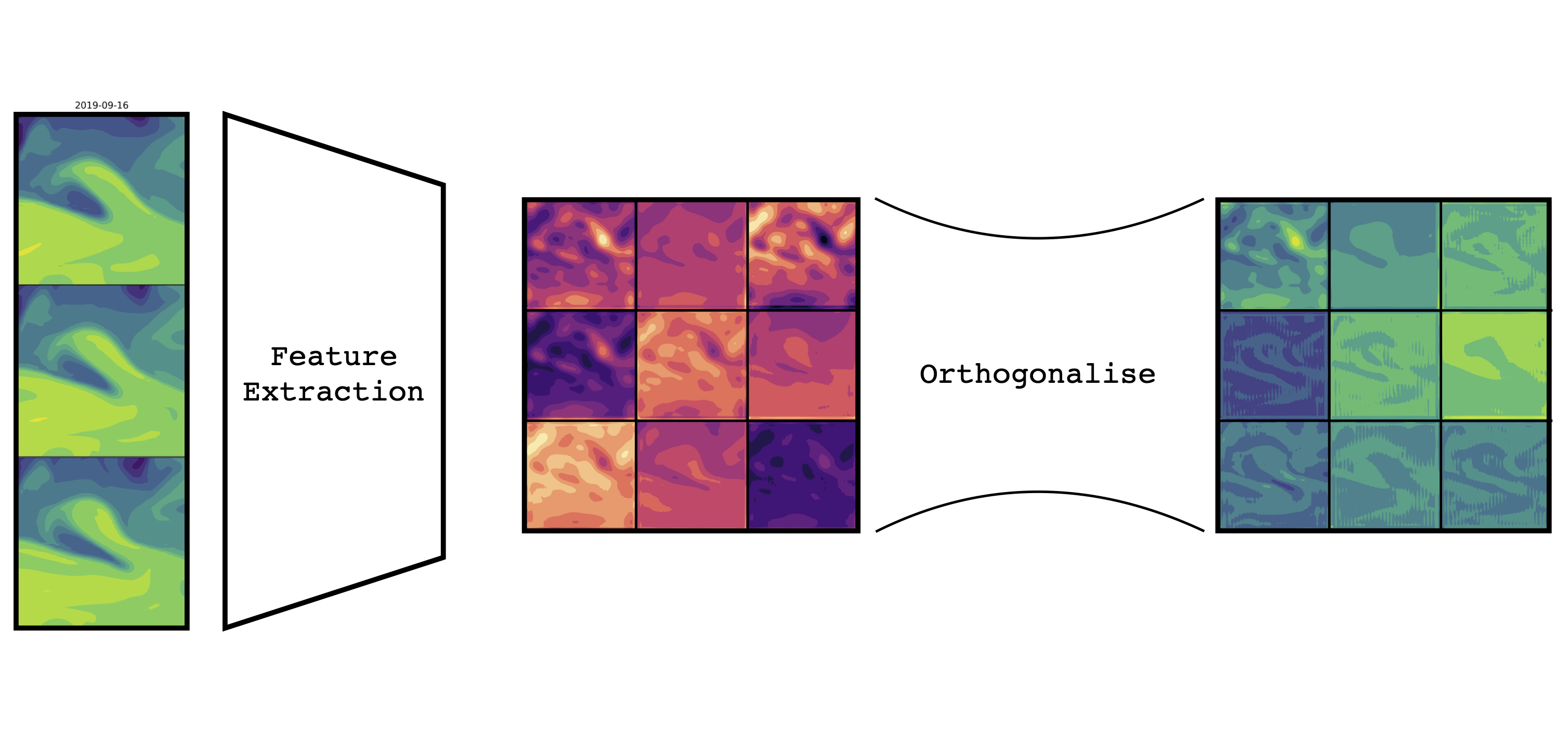}
        \caption{$\texttt{GalerkinBasis}$ feature extraction}
    \end{subfigure}%
    \hfill
    \begin{subfigure}[t]{0.35\textwidth}
        \centering
        \includegraphics[width=.8\linewidth]{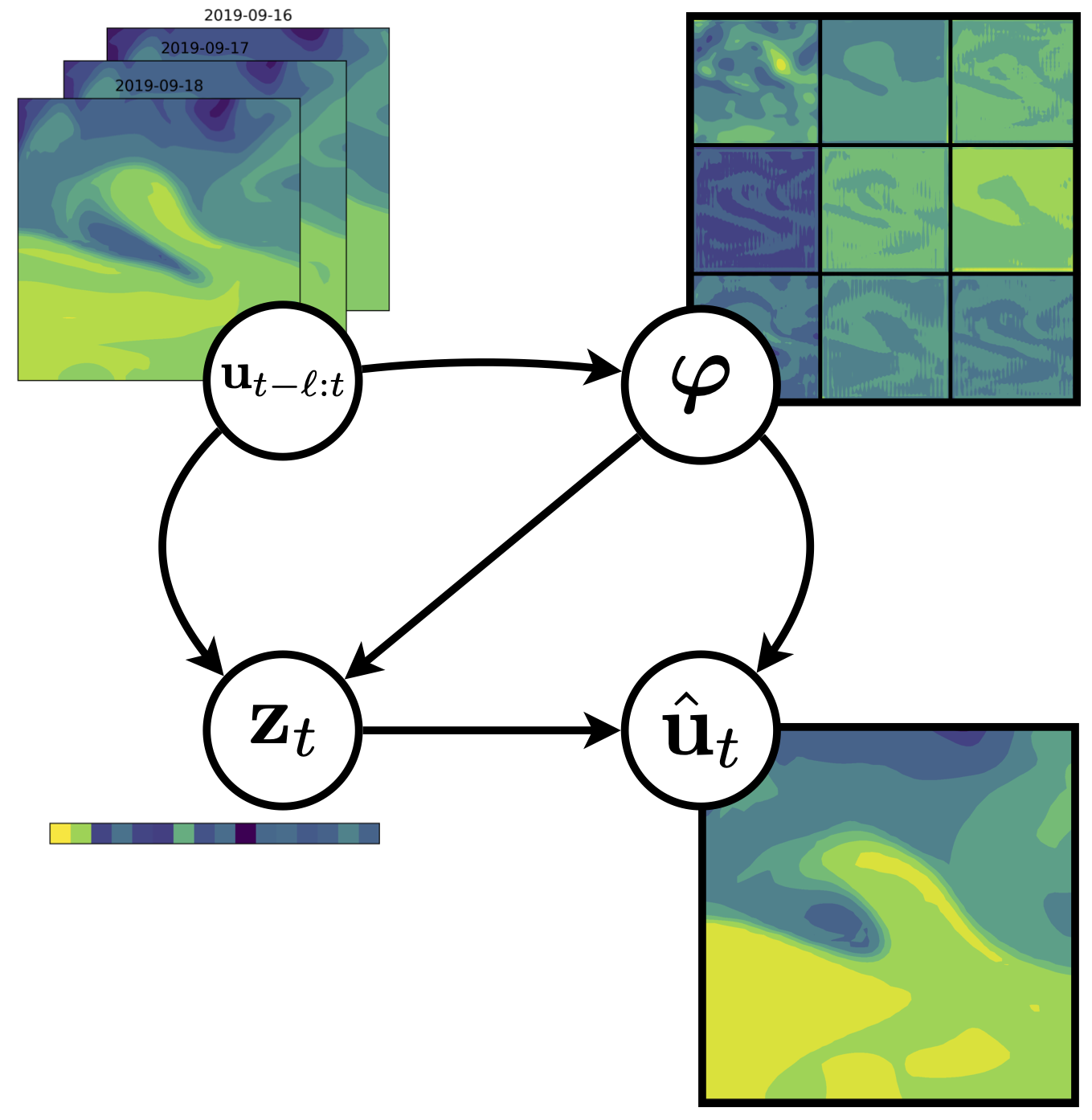}
        \caption{Reconstruction from basis features}
    \end{subfigure}
    \caption{(a) A local input sequence $\{\bu_{k}\}_{t-\ell}^t$
    is passed through the $\GalerkinBasisModel$ feature extraction, 
    to produce a collection of $M$ orthogonal images $\{ \boldsymbol{\varphi}_m \}_{m=1}^M$. (b) An image can be
    projected onto the subspace $\mathrm{span}\{\boldsymbol{\varphi}_m\}$
    and then reconstructed as the linear combination
    $\hat{\bu}_t = \sum_{m=1}^M \langle \mathbf{u}_t, \boldsymbol{\varphi}_m\rangle \boldsymbol{\varphi}_m$
    .
    }
    \label{fig:galerkin_proj}
\end{figure}

In order to project the inputs onto a reduced space it is required to 
construct a set of functions, $\{\varphi_j\}_{j=1}^{M}$, such that 
$\langle \varphi_i, \varphi_j \rangle_{L^2(\Omega)} = \delta_{ij}$,
again we shall work with the vectorised images of these functions and replace
the $L^{2}(\Omega)$ orthogonality condition with a quadrature approximation. Furthermore,
and without loss of generality, we can re-scale the domain to have unit volume so
that we seek a collection of vectors such that 
$\langle \bphi_i, \bphi_j \rangle = \delta_{ij}$, 
$\forall i,j=1,\ldots, M$ under the Euclidean inner
product on $\mathbb{R}^{n_x \times n_y}$.

Any given input $\mathbf{u}_t$ can then projected onto an $M$ dimensional
space by the Galerkin projection 
$\Pi_{\text{Galerkin}}\; : \; \mathbb{R}^{n_x \times n_y} \rightarrow \mathbb{R}^M$
where $\bz_t = \Pi_{\text{Galerkin}} \bu_t$ is the $M$-vector with coefficients
$(\bz)_j = \langle \bu, \bphi_j \rangle$, for $j=1,\ldots, M$. 
While the POD method discussed in the previous section constructs this basis vector
using the \emph{complete} set of inputs, we wish to perform
our projection using only the most recent $\ell$ inputs, relying on the
power of DL approaches to extract sufficient information from 
this reduced set. Our chosen feature extractor will be the 
ubiquitous U-Net \citep{unet}, owing to its demonstrated successes 
in performing feature extraction from image data. Since we seek 
$M$ basis functions we shall consider architectures which takes 
as input an image sequence of shape $(n_x, n_y, \ell)$, and outputs 
an image sequence of shape $(n_x, n_y, M)$. Our 
$\GalerkinBasisModel$ model is then a composition of
this map, with an orthogonalisation of the outputs and we write
\begin{align}
    \{ \bphi_{j}^{(t)} \}_{j=1}^{M} = \GalerkinBasisModel(\bu_{t-\ell}, \ldots, \bu_{t})
    = \texttt{Orthogonalise} \circ \texttt{Unet}(\bu_{t-\ell}, \ldots, \bu_{t}).
\end{align}
Functionally, we have constructed the basis of vectors 
$\bphi^{(t)}_j = \bphi^{(t)}_j(\bu_{u_{t-\ell}}, \ldots, \bu_{t})$, 
which depend only on temporally local values, an overview of the 
composed process is represented in Fig. \ref{fig:galerkin_proj}.

\subsection*{Linear latent convection model}
\label{sec:convec}
The $\GalerkinBasisModel$ provides our desired nonlinear transformation 
of the input sequence onto a linear space parameterised by 
$\bz \in \mathbb{R}^M$, using only information up to time 
$t$. For the remainder of the prediction horizon we 
wish to forward solve the problem using the physically 
informed convection dynamics on a linear space determined by 
the extracted features. We shall assume that 
in this space the dynamics are reasonably well described by 
the linear PDE \eqref{eq:lin_transport}, and use our 
learned features to carry out a projection of the dynamics 
onto a local processes obtained via \eqref{eq:disc_lin_ode}.

\begin{figure}[t!]
    \centering
    \begin{subfigure}[t]{0.47\textwidth}
        \centering
        \includegraphics[width=\linewidth]{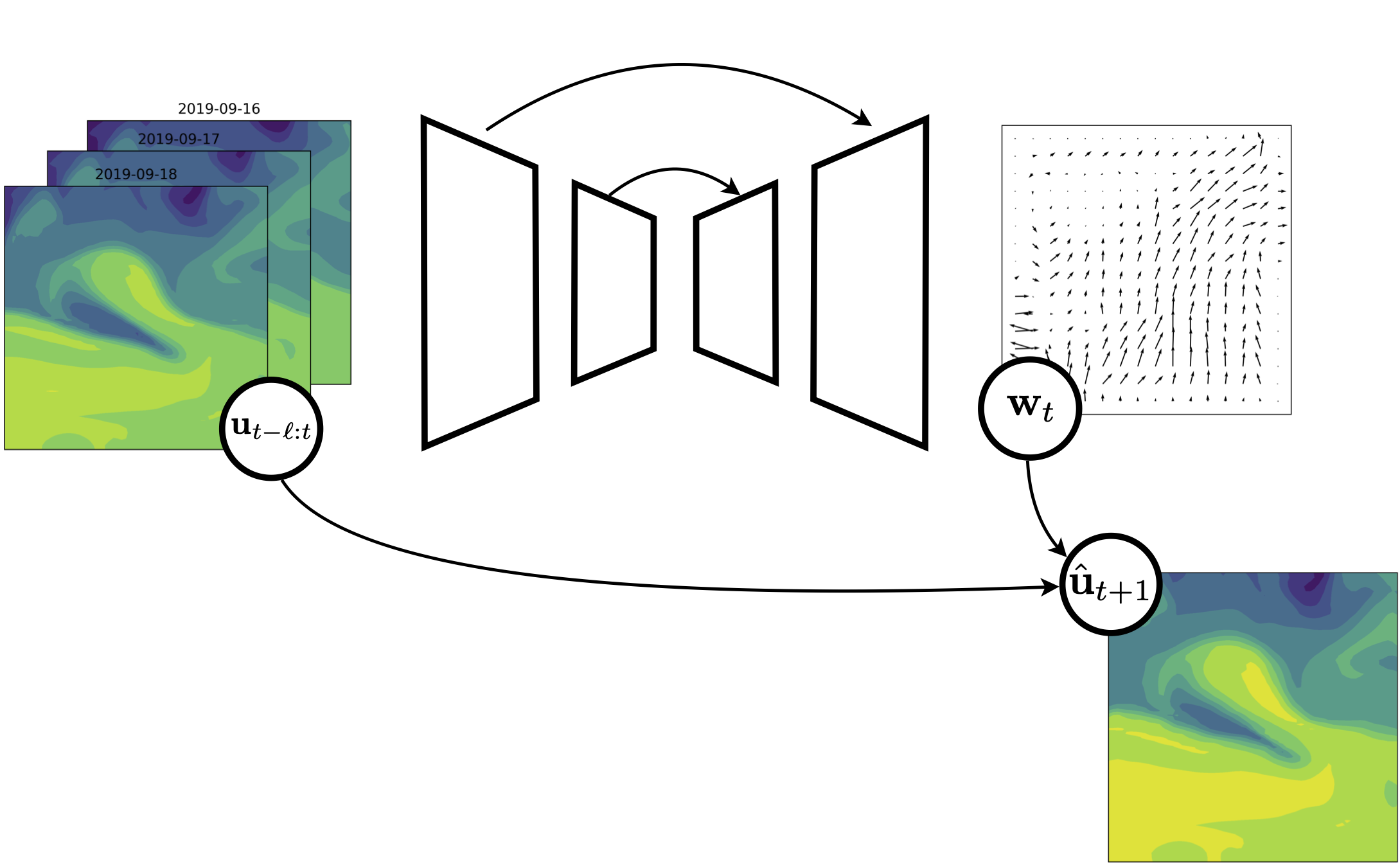}
        \caption{Single time step of \citep{de2018deep}}
        \label{fig:convection_advance_a}
    \end{subfigure}%
    \hfill
    \begin{subfigure}[t]{0.47\textwidth}
        \centering
        \includegraphics[width=\linewidth]{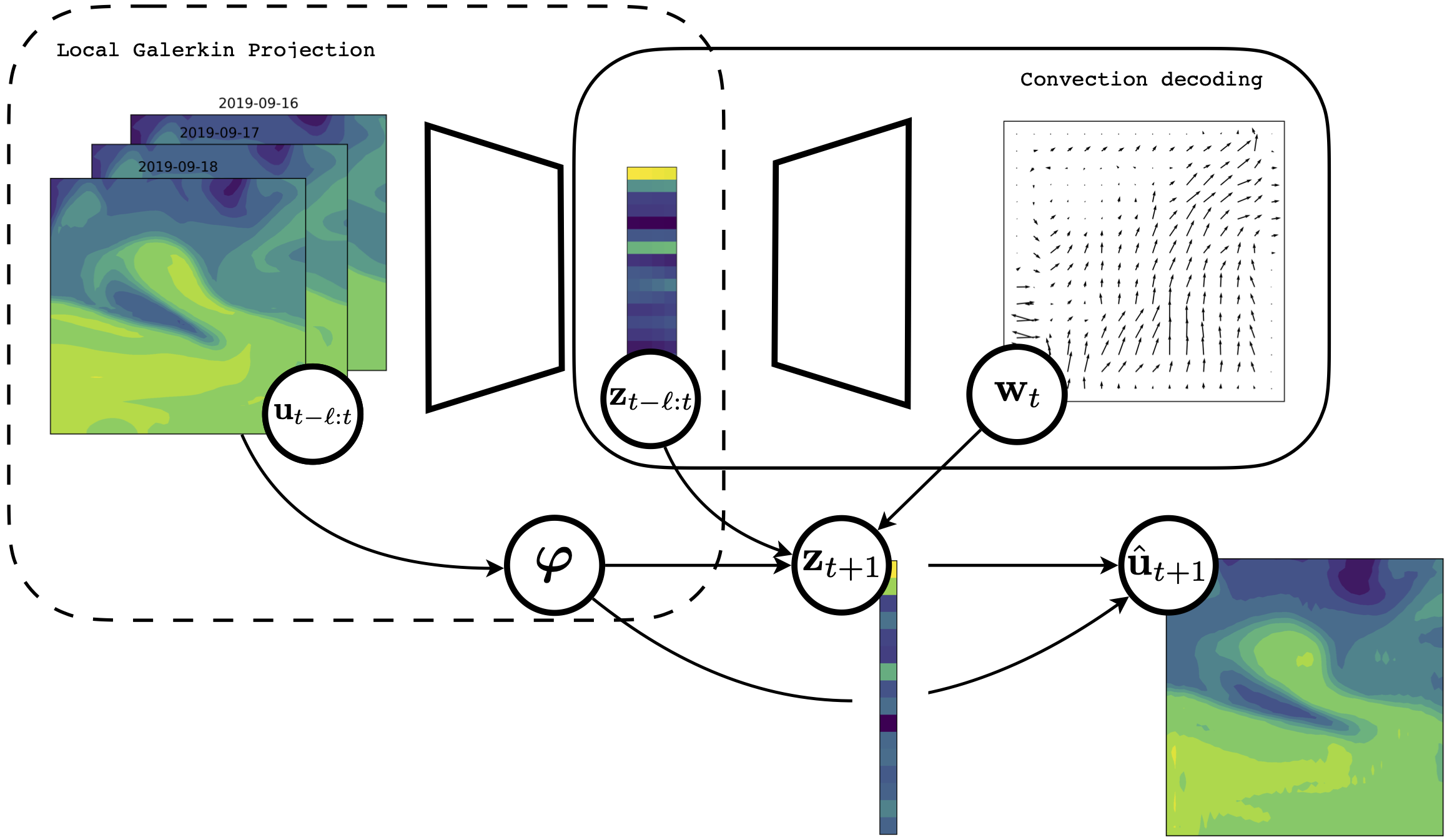}
        \caption{Single time-step of our $\MyModel$ model}
        \label{fig:convection_advance_b}
    \end{subfigure}
    \caption{(a) The method of \citep{de2018deep} uses DL to estimate a
    motion field and then transport the solution in the original data-space,
    no further use is made of the latent representations.
    (b) Our method uses the $\GalerkinBasisModel$ network to encode the inputs 
    to a latent space, and then advances the solution in the latent space using 
    only information from the encoded variables to parameterise the \texttt{Convection} model, before finally 
    reconstructing in the data-space}
    \label{fig:galerkin_advance}
\end{figure}

It remains to parameterise the dynamics with a 
diffusion coefficient, and a transport vector field. We shall refer
to the model component performing this parameterisation as the
$\ConvectionModel$ model, and we shall also allow it to be 
informed by the previous $\ell$ observations. Once this component 
has been specified we shall advance the latent state according to 
an updating scheme of the form
\begin{subequations}
    \begin{align}
        &\alpha^{(k)}, \mathbf{w}^{(k)}= \ConvectionModel (\mathbf{z}_{k-\ell}, \ldots, \mathbf{z}_{k}), 
        \label{eq:convection_advance_a} \\
        &\mathbf{L}_{\varphi}^{(k)} = \texttt{Assembly}(\alpha^{(k}), \mathbf{w}^{(k)},
        \{ \varphi_m(\mathbf{u}_{t-\ell}, \ldots, \mathbf{u}_{t})
        \}_{m=1}^M ) \label{eq:convection_advance_b} \\
        &\mathbf{z}_{k+1} = \bz_{k} + \int_{t_k}^{t_{k+1}}
        \mathbf{L}_{\varphi} \bz_\tau d \tau 
        \label{eq:convection_advance_c}
    \end{align}
\end{subequations} \label{eq:convection_advance}
for $k=0,\ldots, T-1$. First the previous values of the latent process 
are used to build new parameters of the transport model 
\eqref{eq:convection_advance_a}, these are then combined with the features 
to assemble the state dynamic matrix \eqref{eq:convection_advance_b}, and 
finally the linear ODE \eqref{eq:disc_lin_ode} is solved to give the new 
state, \eqref{eq:convection_advance_c}, and this process is repeated until a 
complete set of $T$ latent states are obtained. Our general approach to 
parameterisation of the transporting physical 
model is therefore similar to that in \citep{de2018deep}, but 
crucially our approach propagates the dynamics in the 
\emph{latent space} by way of the Galerkin projection, rather than
applied to the complete surface. This leads to an important difference
since any surface now has a 
finite dimensional representation by $\bz_t$ our convection step
\eqref{eq:convection_advance_a} may be taken as a function of the 
low dimensional latent states, that is we can consider
a decoder which takes the $\ell \times M$ lagged latent states
and outputs the diffusion and transport field over $\Omega$ this is
the top-right block of Fig. \ref{fig:convection_advance_b}. 

Conversely, the approach taken in \citep{de2018deep} requires one to first encode the
surfaces, and then decode them again to estimate the motion field, see Fig.
\ref{fig:convection_advance_a}. While this step necessarily creates an implicit 
latent representation as a byproduct of this
encode/decode step, this representation has no further role to play in 
the model dynamics. 
In contrast, since we have already performed encoding through the Galerkin 
projection, we require no further compression allowing the latent 
representations produced by our approach to feature more directly 
in the dynamical description, see Fig. \ref{fig:convection_advance_b}. 

In summary, our model first uses the output of the $\GalerkinBasisModel$ to
construct a set of orthogonal feature images and then uses the 
Ritz-Galerkin method to project the dynamics onto this space. The 
encoded variables are then passed to the \texttt{Convection} model 
to determine the dynamics, but notably while the convection model 
in Fig. \ref{fig:convection_advance_a} must account for
all of the transformation applied to an input image, our model is able to share the
dynamical description with the $\bphi$-features through the projection
\eqref{eq:disc_lin_ode}. This allows for a more parsimonious
parameterisation of the the \texttt{Convection} model, and since only this smaller
model is called repeatedly during the loop 
procedure \eqref{eq:convection_advance} we achieve a more memory efficient 
implementation. Recognising the partition of our method into 
an orthogonal decomposition, and then a convection informed forward integration, we shall refer to it
as a 
Deep Orthogonal Decomposition for Convection model, or simply $\MyModel$.
\subsection*{Reconstruction and supervised loss}
Finally, having created a full set of latent states 
$\{ \bz^{(k)} \}_{k=t-\ell}^{k=T+t}$, we may use our 
Galerkin features to reconstruct the process. Note that
we have both the historical states,
$\{ \bz_{k} \}_{k=t-\ell}^t$, and the future states
$\{ \bz_{k} \}_{k=t+1}^T$, we also further recall from our discussion
of the POD method in Section \ref{sec:pod} that when combined with the
Galerkin features, the historical states should do a good job of 
reconstructing the inputs, but that the POD method produces the optimal
such reconstruction. Together this partition into historical
and future states suggests the following decomposition of our supervised loss
function
\begin{subequations}
    \begin{align}
        &\mathcal{L}_t = \sum_{k=1}^{T} \rho\left( u_{t+k}
        - \sum_{m} (\bz_{t+k})_{m} \varphi_m (\bx) \right)
        + \lambda_{\text{div}}\| \nabla \cdot \mathbf{w} \|^2 \\
        &\qquad + \lambda_{\text{recon}} \sum_{k=1}^{\ell} \rho\left(u_{t-k} - 
        \sum_{m} (\bz_{t-k})_{m} \varphi_m (\bx) \right).
        \label{eq:supervised_loss}
    \end{align}
\end{subequations}
The first two terms of \eqref{eq:supervised_loss} are analogous to those
in \eqref{eq:their_supervised_loss}, playing a similar role with regards to
measuring the accuracy of the prediction through a metric $\rho$, 
while regularising the divergence of the output $\bw$ of the 
$\ConvectionModel$ to ensure a valid divergence free vector field.

The final term is specific to our method, and addresses the
optimality condition of the POD basis as discussed in Section \ref{sec:pod}, 
in particular we commented that when used to reconstruct 
an ensemble the POD basis gives the minimal reconstruction error 
for a particular dimension of the subspace. The coefficient
$\lambda_{\text{recon}}$ therefore controls the extent to which our
local basis decomposition
attempts to recover a version of this optimally condition. This trade-off
helps ensure that the $\GalerkinBasisModel$ continues to have an identifiable
role as an encoder/decoder of the surfaces given its additional role in 
assembling the discretised differential operator $\mathbf{L}_{\varphi}$. This dual
role of the $\bphi$-features is one of the principal strengths of 
our method, allowing dynamical
components lost by the reduction to the overly simplistic 
PDE \eqref{eq:lin_transport}, to be recaptured by the DL-features, at the
expense of the optimality of these same features as re-constructors of the
surface.
\section{Experiments}
\label{sec:experiments}
We now apply our method to the large-scale convective systems encountered
in the climate sciences in order to assess
the accuracy and computational efficiency of our method compared to alternatives.
All experiments were conducted on a single Nvidia GeForce RTX 2060 SUPER GPu 
with 8GB of memory. For additional experiments and visualisations see 
the supplementary material.

\subsection{Convective seasurface temperature}
\begin{wrapfigure}{L}{0.5\textwidth}
\centering
\includegraphics[width=0.8\linewidth]{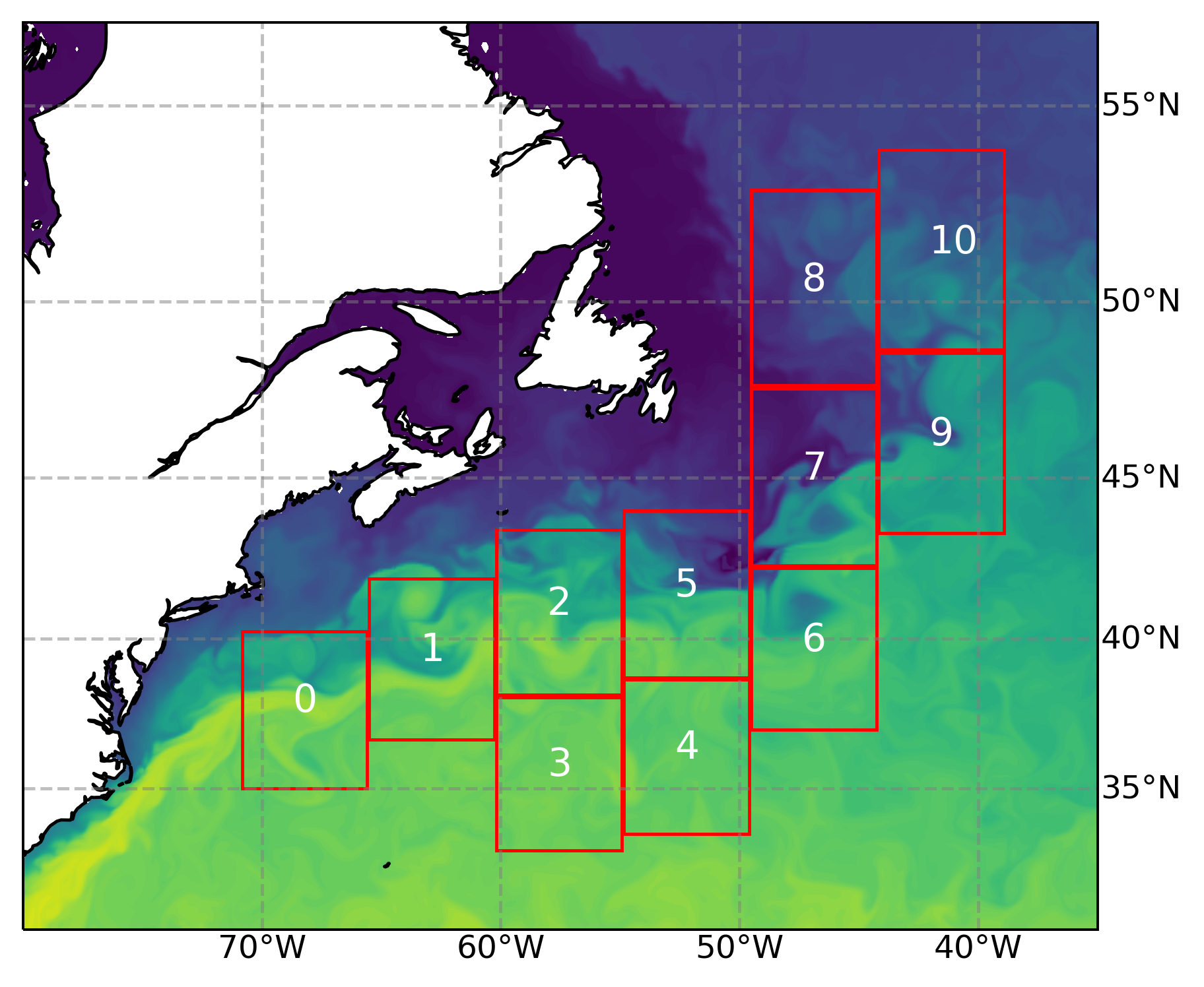}
\caption{Regions used for the sea surface temperature experiment}
\label{fig:sst_regions}
\end{wrapfigure}
As in \citep{de2018deep} we demonstrate our model on the SST data obtained 
from the NEMO (Nucleus for European Modelling of the Ocean) engine 
\citep{madec}. We extract the 64 by 64 pixel sub-regions shown in Fig. 
\ref{fig:sst_regions}, and use data from 2016-2018 to train the model, a
give a total of 11935 acquisitions. We then test on data from the years 2019 
using the same regions giving 3894 test instances. The regions were selected 
to capture the complicated nonlinear mixing that occurs as warm-water is transported 
along the gulf stream and meets the colder water of the arctic regions. We 
assume that on the short-term
horizons we are interested in, the information within a region is sufficient
for prediction and do not incorporate information from adjacent regions. 
Data is normalised to remove trends on a yearly 
timescale by taking the mean and standard deviations over the day of the
year.

We compare the method we have introduced to the physics informed model
of \citep{de2018deep}, as well as a convolutional LSTM (Conv-LSTM) introduced by
\citep{shi, shi15} which enhances LSTM networks with 
a spatial convolution to better model spatio-temporal process, and finally
to compare with ``phyiscs-free'' approaches we use a U-Net as proposed in 
\citep{AYZEL2019186, lebedev, google} which treats
prediction as an images-to-images problem, and makes no attempt to embed
recurrent dynamical or physical structure. All models were
implemented in Tensorflow \citep{tensorflow2015-whitepaper}, apart from \citep{de2018deep} for which we use publicly available code.
\footnote{A PyTorch \citep{torch} implementation for the model 
of \citep{de2018deep} is available at \url{https://github.com/emited/flow}}

\begin{table}[ht]
\caption{Comparison of methods on the SST data. Average score is the
mean squared error (MSE), No. of parameters is the total number of trainable
parameters in each model, and run-time is the mean time per-batch with the 
maximum batch size that can fit in memory. We fit
our model with $M=16$ $\bphi$-features}
\label{tab:sst_results}
\vskip 0.15in
\begin{center}
\begin{small}
\begin{sc}
\begin{tabular}{lcrr}
\toprule
& Average Score (MSE) & No. of Parameters & Run-Time [s] \\
\midrule
Conv-LSTM\citep{shi15} & 0.2179 & 1,779,073 & 0.43  \\
U-Net \citep{lebedev, google} & 0.1452 & 31,032,446 & 0.79\\
Flow \citep{de2018deep} &  0.1344 & 22,197,736 & 0.60 \\
\texttt{DOD-Convec} & $\mathbf{0.1242}$ & 10,106,339 & 0.48\\
\bottomrule
\end{tabular}
\end{sc}
\end{small}
\end{center}
\vskip -0.1in
\end{table}

Results are displayed in Table \ref{tab:sst_results}, examining the 
test error for each method we see that our method demonstrates 
superior performance with a lower test error than the purely data-driven approaches and the more dynamic alternatives. As discussed in Sec. 
\ref{sec:convec} our approach leads to a more parsimonious parameterisation 
of the convection field than \citep{de2018deep}.
With regards to run-time the U-Net model is faster, and more 
memory efficient than the alternatives,
including our own, which make use of a recurrent 
structure to advance the predictions. 
In spite of this our method still demonstrates competitive run time, with
the increased accuracy a more than acceptable trade-off.

\begin{figure}
    \centering
    \includegraphics[width=\textwidth]{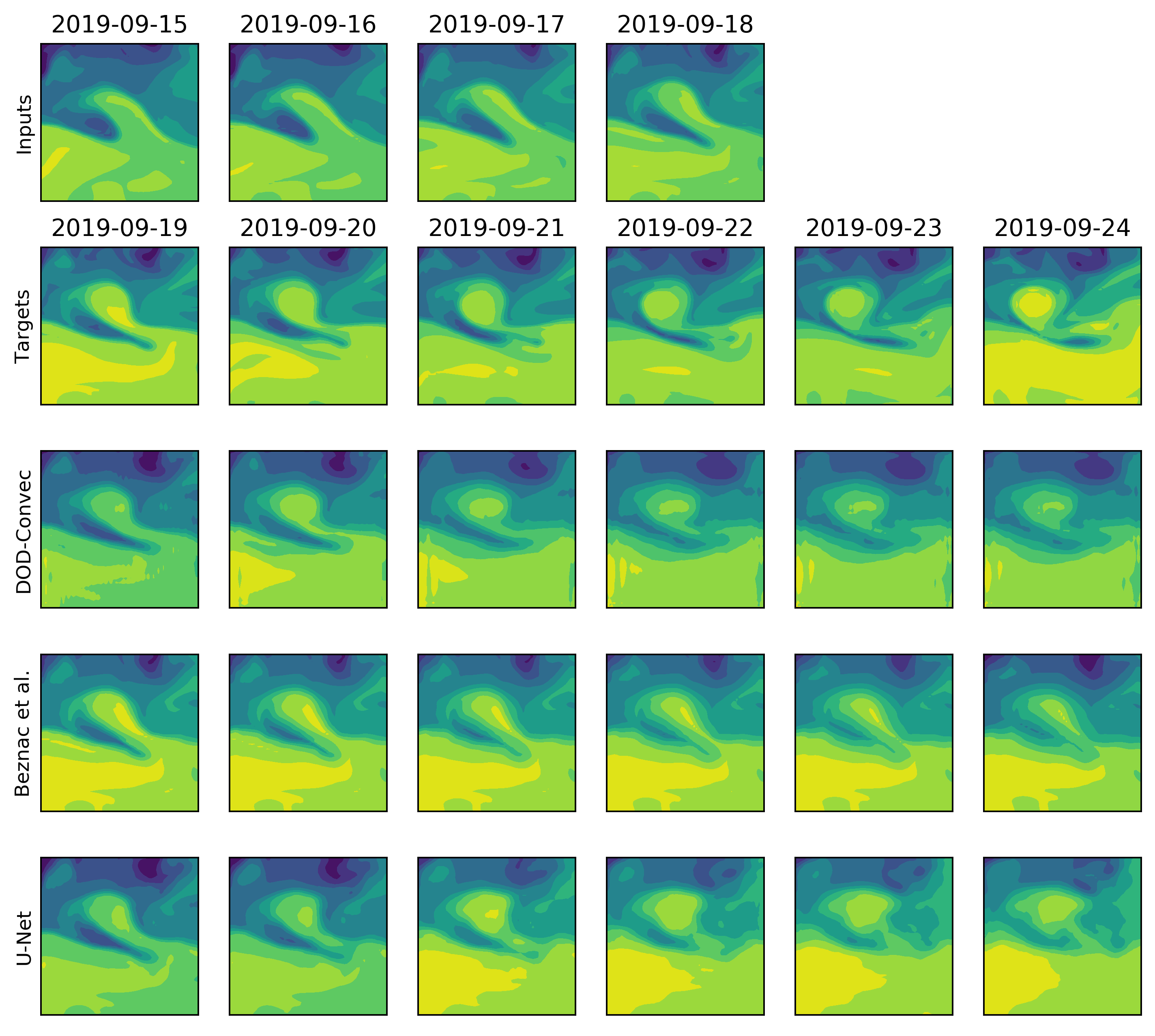}
    \caption{Predicted surfaces from an input sequence of length four, top row,
    compared to a target output sequence of length six, second row.
    from a test set sequence in Region 2 of Fig. \ref{fig:sst_regions}}
    \label{fig:prediction}
\end{figure}

In Fig. \ref{fig:prediction} we plot a representative sequence from the
test set, in which we observe that our method, row three, seems to be better at
identifying the ``whorl'' pattern formations that characterise turbulence,
but that the U-Net feature extraction also does a job job of identifying these
features, but only on a limit time-horizon with the structure degrading
as the sequence continues, this is most noticeable in the loss
of structure of the in-flowing cold flux in the top-right of the images 
in row four. On the other-hand the method of \citep{de2018deep} does
a better job capturing and preserving linear features, this is likely 
because this method is ultimately solving
a linear PDE \eqref{eq:raleigh_benard_a}, and identifying a 
convection model from data alone that
will do the ``violence'' \citep{berkooz} of a nonlinear model from data 
alone is hard. By projecting to a latent space, and allowing this
linear space to be adaptivly determined by nonlinear transformations 
of the local inputs we are better able to recover nonlinear features 
with a simpler convection model, and while we do pay a visible price in 
the smoothness of our reconstructions, overall we achieve better performance as presented in Tab. \ref{tab:sst_results}.

\section{Related work}

\textbf{Spatio-temporal dynamical modelling} Traditional 
dynamical modelling approaches have been based on either 
optical flows \citep{fleet, woo_optical_rainflow} 
which estiamte the transport vector field by examining the
differenced input sequence, or else numerical models 
\citep{bereziat:hal-01245369} where the transporting field 
emerges as the output of a calibrated physical model. More
principled physical models have the ability to display 
complex long-range spatio-temporal dependencies, but deriving
and then calibrating such models is significantly harder than
the DL and hybrid methods we consider.

\textbf{Physics free deep learning} As noted by 
\citep{google} ``Physics free" deep learning approaches can be 
boradly divided into attempts to model the time evolution as
a temporal sequence, or else those which treat the problem as
learning an unknown ``image(s)-to-images'' map. The former
include the convolutional LSTM model of \citep{shi, shi15}, 
as well as video generation methods \citep{DBLP:journals/corr/MathieuCL15, vondrick}
For the latter option U-Net based approaches have been popular, \citep{AYZEL2019186, lebedev} and \citep{google} who argue that
the overly simplistic assumptions in optical flow or poorly specified
numerical models are often unviable, so that free-form DL methods 
display superior performance methods tied to a misspecified physical model.
While our work does use an overly simplistic model of the dynamics, by
projecting on the latent space we are able to ensure this 
simplification does not overly constrain the resulting predictions.

\textbf{Physics informed deep learning} Aside from the
work of \citep{de2018deep} which we have discussed in Section
\ref{sec:background}, we also mention the more complicated U-Net 
architecture inspired by multiscale approximations to
the Navier-Stokes equations in \citep{wang2020towards}, since
this model uses physical information to design
a specific U-Net architecture there some overlap with
the already mentioned images-to-images methods.
Finally we note that we have advanced the solution to the problem
by solving something which closely approximates the solution of a PDE.
Alternative approaches \citep{Raissi, Berg, zhu} 
have avoided this explicit solve step
and instead use flexible DL surrogates of the output map, with the PDE 
operators re-appearing as regularisers. This is a much more dramatic 
regularisation than the $\lambda_{\text{div}}$ loss we have considered, 
and we are unaware of this complete operator regularisation scheme having 
been applied to convective non-linear systems at the scale considered in 
the experiments.

\section{Discussion}
In this work we have combined the powerful feature extraction capabilities
of DL, with minimal physical representations of the dynamics to 
introduce a physically informed model demonstrating superior predictive
performance on the convective physical processes that dominate the
atmospheric and fluid transport problems pervading our natural world.
To the extent that this model is ``physically informed'' we have a 
priori specified only a minimal representation of the dynamics, we
justify this by our desire to avoid overly strong, and so impossible to
justify, assumptions on the complex generating mechanism and therefore
maintain as much as possible the ability of the model to learn
the underlying dynamics from data. Crucial to this has been our efforts
to ensure that the latent representations formed by DL-encoding are more strongly
entwined with the dynamics of the process than previous approaches, 
leading to a model that demonstrates superior predictive performance,
and improved recovery of temporally persistent nonlinear features.
\subsection*{Broader Impact}
Climate change manifests not only as long-term trends, but also
as the increased short-term volatility of these processes, and as 
such the development of near-term prediction methods which can 
be rapidly evaluated and re-calibrated is only increasing in 
importance. Accurate near term prediction of rapidly
changing events is vital for public safety in situations
including extreme flooding, cyclone formation, and forest fires 
-- all of these systems possessing the informative transport 
features which would allow our method to be successfully applied.
These applications will continue to add to the evidence we have
already provided that physical principles still have an important 
role to play in designing models, and argue against the ``physics
free'' philosophy. We demonstrate that there is much to be gained 
from strongly connecting learned latent representations with global
model dynamics, achieving not only interpret-able latent spaces,
but also interpretable dynamics on these spaces. The resulting 
models demonstrate better performance, and
have a clearer physical picture, and we envision our work continuing
to invite methodological study of the generalisation benefits to
be obtained by this stronger entwining between encoded representations
and the original data-space dynamics.

Despite the progress we have made, predicting these systems remains a
hard problem, and it is unlikely that any one method will work best in
all scenarios and so appropriate aggregation strategies should be adopted.
However, our discussion of the sea surface predictions in 
Sec. \ref{sec:experiments} demonstrates that our model does a better 
job of predicting certain nonlinear patterns in convective flows, and so will
be an effective complement to those existing methods which 
perform well on more linear flows. Finally while we have presented an effective
supervised learning method, it will be important to extend the work we have 
done to a generative probabilistic framework to provide further uncertainty 
quantification.

\small

\bibliography{references}

\appendix
\section*{Appendices}
\addcontentsline{toc}{section}{Appendices}
\renewcommand{\thesubsection}{\Alph{subsection}}

\section{Ritz-Galerkin method for PDEs}
\label{supp:sec:ritz-galerkin}
Since the forward map is unavailable in closed form it becomes necessary to solve 
the PDE numerically. To accompany the necessarily brief review in the main paper 
we provide an extended presentation of how to discretise a PDE on the model 
Poisson problem, though see also \citep{reddy} for more information

\begin{subequations}
\begin{align}
    -\Delta u(\bx) &= f(\bx)  \quad \mbox{ for } \bx \in \Omega \label{eq:poisson_a} \\
    u(\bx) &= g(\bx) \quad \mbox{on } \partial \Omega.
\end{align}  \label{eq:poisson}
\end{subequations}

To discretise one first multiplies \eqref{eq:poisson_a} by some test function $v \in \hat{V}$ and then performs an integration by parts, it is standard to further assume that the test functions vanish on the boundary, so one arrives at the classical form of the \emph{variational problem}: 
find $u \in V$ such that
\begin{align}
    \int_{\Omega} \nabla u(\bx) \cdot \nabla v(\bx) d\bx = 
    \int_{\Omega} f(\bx) v(\bx) d\bx - 
    \int_{\partial \Omega} g(\mathbf{s}) v(\mathbf{s}) ds \label{eq:finite_var_problem}
\end{align}
for any test function $v \in \hat{V}$, this is the \emph{weak form} of the
classical problem \eqref{eq:poisson}.

To numerically implement this idea one replaces the trial and test spaces by
finite dimensional sub-spaces $V_h$ and $\hat{V}_h$ respectively, with corresponding
basis functions denoted by $\phi_i$, and $\hat{\phi}_i$.

Once we have specified this basis we can then search for solutions with a representation
of the form $u(\bx) = \sum_{j=1}^M (\boldsymbol{\xi})^{u}_j \phi_j(\bx)$, where 
$\boldsymbol{\xi}^u \in \mathbb{R}^{M}$ is the vector of coefficients 
parameterising the finite-dimensional representation of $u$. Inserting this representation into \eqref{eq:finite_var_problem} we arrive at the 
finite-dimensional version on the weak-form given by the system of equations

\begin{align*}
    &\sum_{j=1}^N (\boldsymbol{\xi}^u)_j \int_{\Omega} \nabla \phi_j(\bx) \cdot \nabla \hat\phi_i(\bx) d\bx = 
    \int_{\Omega} f(\bx) \hat{\phi}_i(\bx)d\bx 
    - \int_{\partial\Omega} g(\mathbf{x}) \hat{\phi}_i(\mathbf{s}) d\mathbf{s},
\end{align*}

which must hold for each basis vector $\hat{\phi_i}, i=1,\ldots, M$. We may 
represent this in equivalent matrix-vector notation as 
$\mathbf{A} \boldsymbol{\xi}^u = \bff$, where the \emph{stiffness matrix} 
and \emph{load vector} are given by

\begin{subequations}
\begin{align}
    (\mathbf{A})_{ij} &= \int_{\Omega} \nabla \phi_j(\bx) \cdot \nabla \hat{\phi}_i(\bx) dx 
    \label{eq:stiffness_matrix} \\
    (\bff)_i &= \int_{\Omega} f(\bx) \hat{\phi}_i(\bx) dx 
    - \int_{\partial\Omega} g(\mathbf{s}) \hat{\phi_i}(\mathbf{s}) d\mathbf{s}.
    \label{eq:load_vector}
\end{align}
\end{subequations}

Following the same procedure we can arrive at a weak-form discretisation of 
the transport equation \eqref{eq:lin_transport}. In this case the discrete 
representation analogous to \eqref{eq:stiffness_matrix} is given by 
$\mathbf{L}[\bz]$ where

\begin{align}
&(\mathbf{L}[\mathbf{z}])_{ij} = \int_{\Omega} a(\bx) \nabla \phi_j(\bx)
\cdot \nabla \hat{\phi}_i(\bx) dx 
+ \int_{\Omega} \mathbf{w}(\bx) \cdot \nabla \phi_j(\bx) 
\hat{\phi}_i (\bx)dx. \label{eq:transport_assembly}
\end{align}

\section{Network architectures}

\subsubsection*{U-Net architectures}
We use a U-Net \citep{unet} for both our basis feature extractor, and in its
own right as a benchmark DL method. These take the standard U-Ne form being
divided into an encoder and a decoder, the encoder is a sequence of computational
units called the \emph{Downsample Blocks} and the encoder a sequence of 
\emph{Upsample Blocks}, long range skips connections are then provided to connect
each downsample block with the corresponding upsampling block in the decoding
phase. The basis form of these blocks is given by
\begin{itemize}
    \item Downsample Block: $\text{MaxPooling} \rightarrow
    \text{Conv2D} \rightarrow \text{ReLU} \rightarrow \text{Conv2D} \rightarrow \text{ReLU}$
    \item Upsample Block: $\text{Upsample} \rightarrow 
    \text{Conv2D} \rightarrow
    \text{Relu} \rightarrow
    \text{Conv2D} \rightarrow
    \text{Relu} \rightarrow
    \text{Conv2D}$
\end{itemize}
where the upsampling operation doubles the pixel resolution in both directions using nearest
neighbour interpolation.

The U-Net used in the experiments has four down-sample and up-sample blocks, the smaller
U-Net used as our feature extractor has three.

\subsubsection*{Convolution decoder}
As discussed in Sec. \ref{sec:methodology} our method estimates the transport vector field
and diffusion coefficient from the latent variables $\mathbf{z} \in \mathbb{R}^{\ell \times M}$,
this is done in two phases, first using a sequence of dense networks, and then convolutions.
For the $64 \times 64$ image patches used in the sea surface temperature experiment this
network takes the form
\begin{itemize}
    \item Dense phase: $
    \text{Flatten} \rightarrow 
    \text{Dense}(64, \text{'relu'}) \rightarrow 
    \text{Dense}(256, \text{'relu'}) \rightarrow 
    \text{Dense}(1024, \text{'relu'}) \rightarrow
    \text{Dense}(1024, \text{'relu'}) \rightarrow
    \text{Dense}(1024, \text{'relu'})
    $
    \item Convolution phase: $
    \text{Reshape}((4, 4, 64)) \rightarrow 
    \text{Upsample} \rightarrow 
    \text{Conv2D}(32, 3, \text{'relu'}) \rightarrow 
    \text{Upsample} \rightarrow 
    \text{Conv2D}(16, 3, \text{'relu'}) \rightarrow 
    \text{Upsample} \rightarrow 
    \text{Conv2D}(8, 3, \text{'relu'}) \rightarrow 
    \text{Upsample} \rightarrow 
    \text{Conv2D}(3, 3)
    $
\end{itemize}
where $\text{Dense}(n, \text{'relu'})$ is a densely connected layer
with $n$ features and ReLU activation function. Similarly
$\text{Conv2D}(n, k, \text{'relu'})$ is a 2D convolution layer
with $n$ features and a $k \times k$-kernel operation. The
Upsample operation is the same as for the U-Net architecture just
described. The Flatten operation takes the sequence of latent
states and flattens them into a single $\ell \times M$ vector.

The output of this as $(64, 64, 3)$ image, final we split on
the last dimension to produce the $(64, 64, 1)$ diffusion
field, and the $(64, 64, 2)$ transport vector field, and
the diffusion field is passed through an additional softplus
layer to ensure positivity.

\section{Further experimental details}
\subsection{{NEMO} Sea Surface Temperature data}
The NEMO (Nucleaus for European Modelling of the Ocean) engine \citep{madec} is
a state-of-the-art framework for modelling the convective processes that drive
the worlds oceans, including temperature, salinty, sea ice area and thickness etc., 
and is provided by the Copernicus Marine Service portal with product identifier 
\texttt{GLOBAL\_ANALYSIS\_FORECAST\_PHY\_001\_024}. While it is a simulation
engine, it accumulates historical data to produce a synthesised estimate of the state
of the system using data assimilation methods, as such the data does follow the true
temperature, making this an ideal dataset to apply our method to. 

Our dataset is formed by first decomposing a larger study region into 64 by 64 pixel
image patches, these are displayed in Figure \ref{fig:sst_regions}. Our regions differ 
from those used in \citep{de2018deep} in that we have selected our grouping to more closely
follow the movement of the gulf stream up the Eastern seaboard of North America
and then accross the Atlantic, mixing with the colder water from the Arctic regions.
This choice allows us to model the most interesting regions, with the complex
nonlinear mixing dynamics we are most interested in. 

Due to a re-calibration of the \texttt{GLOBAL\_ANALYSIS\_FORECAST\_PHY\_001\_024} product
we only have access to data from January 1st, 2016 onwards and so cannot use the
same temporal range as considered in \citep{de2018deep}. Instead we use data
from 2016--2018 as training data, and then the data from 2019 to test on. We also
partitioned our data into length 4 input sequences, and then predicted on the next
8 sequences. Leading to 11935 training examples and 3984 test examples, we also
test our data on the same set of regions as used for training. 

\begin{figure}
    \centering
    \includegraphics[width=\linewidth]{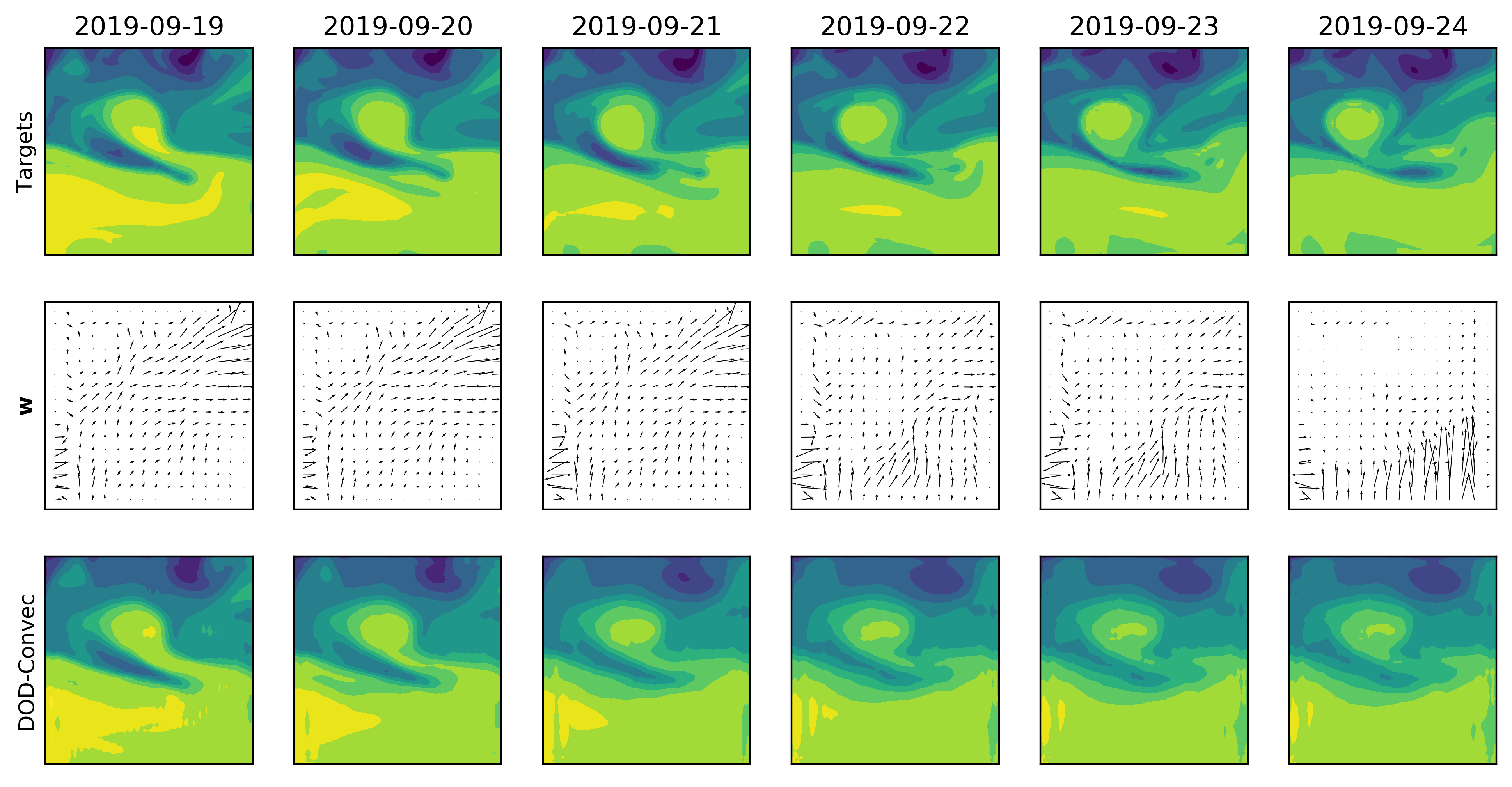}
    \caption{Estimated convection field for our model.}
    \label{fig:my_label}
\end{figure}

\textbf{Additional figures} In Fig. \ref{fig:my_label} we display the 
predictions from our model first presented in Fig \ref{fig:prediction},
but now including also the convection field used to advance our model.
This vector field displays both long range stream lines, as well as localised
regions around which a particle rotates around a center, furthermore the
richness of all of these fields was obtained by decoding a
$4 \times 16$ dimensional vector, and not as an image-to-image transform.
This parsimonious deconvolution from the latent space, combined with the
ability of our method to learn complex vector fields is the principle
component of its demonstrated success on convective fields such as the 
sea surface temperature data.

\subsection{Precipitation nowcasting}
For this experiment we consider the problem of providing short
term precipitation forecasts from Doppler radar images as 
considered in \citep{google}. The data comes from the multi-radar 
multi-sensor (MRMS) system developed by 
NOAA National Severe Storms Laboratory, and provides precipitation
rate updates at 2 minute intervals over a spatial resolution of
1km $\times$ 1km. We use the MRMS dataset \citep{mrms} re-sampled
to 10 minute intervals. The dataset covers the whole of the United States,
but we restrict out attention to sixteen $128 \times 128$ regions in
the Pacific Northwest, and then further restrict our attention to the
two wettest regions in the training set. We train on data from January 2018 and then 
test on the first week of November 2017 giving 8904 training instances and
2568 test instances. Following \citep{google} we frame the problem as a
classification problem and discretise the precipitation data using
three thresholds in units of millimetres of rain per hour, these
ranges are given by $[0, 0.1), [0.1, 1.0), [1.0, 2.5)$ and $[2.5, \infty)$,
and the model is then trained using a cross-entropy loss target these pixel values.

The resulting accuracy on the classification problem is
displayed in Table \ref{tab:precip_results}, which demonstrate
that there is little to choose between the two approaches
in this instance. It would appear that once the process has
been coarsened in this way there is no significant difference 
between the physics free method and our physics informed method
on the precipitation data, the implications of this are the subject of
further study, however it is ultimately reassuring that our method
proves no worse on this problem, and indeed significant better on
the ``full-image'' temperature data reported in Sec. \ref{sec:experiments}

\begin{table}[ht]
\caption{One hour ahead precipitation nowcasting
accuracy.}
\label{tab:precip_results}
\vskip 0.15in
\begin{center}
\begin{small}
\begin{sc}
\begin{tabular}{lcc}
\toprule
& Training Accuracy & Test Accuracy \\
\midrule
U-Net \citep{lebedev, google} & 0.9353 & 0.8047 \\
\texttt{DOD-Convec} & 0.8445 & 0.8052 \\
\bottomrule
\end{tabular}
\end{sc}
\end{small}
\end{center}
\vskip -0.1in
\end{table}

\end{document}